\documentclass[amsmath,amssymb,amsfonts,preprintnumbers,showpacs]{revtex4}
\usepackage[caption=false]{subfig}
\usepackage{graphicx}
\usepackage{bm}
\usepackage{booktabs,array,multirow}
\usepackage[usenames]{color}

\begin{document}

\title{Optical Absorption by Indirect Excitons in a Transition Metal Dichalcogenide Double Layer}
\author{Matthew N. Brunetti$^{1,2}$, Oleg L. Berman$^{1,2}$, and Roman Ya. Kezerashvili$^{1,2}$}
\affiliation{%
$^{1}$Physics Department,  New York City College of Technology\\
The City University of New York,
  300 Jay Street,   Brooklyn NY, 11201, USA \\
$^{2}$The Graduate School and University Center\\
The City University of New York,
New York, NY 10016, USA \\
}

\date{\today}

\begin{abstract}
  We study optical transitions in indirect excitons in transition metal dichalcogenide (TMDC) double layers separated by an integer number of hexagonal boron nitride ($h$-BN) monolayers.
  By solving the Schr\"{o}dinger equation with the Keldysh potential for an indirect exciton, we obtain eigenfunctions and eigenenergies for the ground and excited states and study their dependence on the interlayer separation, controlled by varying the number of $h$-BN monolayers.
  The oscillator strength, optical absorption coefficient, and optical absorption factor, the fraction of incoming photons absorbed in the double layer, are evaluated and studied as a function of the interlayer separation.
  Using input parameters from the existing literature which give the largest and the smallest indirect exciton binding energy, we provide upper and lower bounds on all quantities presented.
\end{abstract}

\maketitle

\section{\label{sec:intro}Introduction}

  Since the experimental realization of highly conductive graphene monolayers in 2004~\cite{novoselov2004electric}, the field of condensed matter physics has seen explosive growth in interest in two-dimensional (2D) materials.
  In recent years, experimental success in isolating stable monolayers of 2D insulators such as hexagonal boron nitride ($h$-BN)~\cite{Dean2010} and 2D semiconductors such as transition metal dichalcogenides (TMDCs)~\cite{novoselov2005two} has, along with graphene, given researchers all of the materials required to build chips capable of powering the electronic devices of the future.
  In addition to their promising potential role as the semiconducting material in ``traditional'' electronic devices~\cite{Novoselov2012,Geim2014}, TMDCs have also garnered considerable interest for their potential role in optoelectronic devices~\cite{Xia2014,Wang2012} due to their exceedingly strong optical absorption and photoluminescence properties~\cite{Yan2014a,Steinhoff2015,Qiu2013}.

  One of the reasons why TMDCs are interesting from an optoelectronics standpoint is their ability to form excitons with very large binding energies.
  An exciton is the bound state of an electron in the conduction band and a hole in the valence band.
  Excitons in TMDC monolayers are referred to as direct excitons, to distinguish them from indirect excitons, which are formed in a TMDC double layer when an electron in one layer becomes bound to a hole in another, parallel layer.
  %Indirect excitons may be formed in homogeneous TMDC double layers by applying a voltage perpendicular to the plane of the TMDC monolayers.
  %In heterogeneous TMDC double layers with no dielectric, also known as TMDC heterostructures, the mis-alignment of the band gaps of the two TMDC monolayers causes a type-II heterojunction in the band structure of the double layer \textendash{} this causes indirect excitons to form spontaneously when a direct exciton is first created in either of the two TMDC monolayers.
  %When indirect excitons are formed in TMDC heterostructures with no dielectric, they are referred to as charge transfer excitons because of their spontaneous formation.

  In a field such as optoelectronics, it is essential to have a thorough understanding of not only the optical properties that lead to the formation of excitons, but also the optical properties of these excitons themselves, namely the oscillator strengths, absorption coefficients, and transition energies to the excitonic excited states.
  In addition, theoretical predictions of the aforementioned optical quantities are an essential tool for interpreting experimental results.
  % By providing a complete set of optical quantities for experimentally feasible setups characterized by a TMDC double layer separated by an integer number of $h$-BN monolayers, experimentalists can verify the formation of indirect excitons by checking for the presence of the characteristic quantities presented here such as optical transition energies and absorption coefficients.

  To this end, there has been a significant amount of experimental work dedicated to understanding the electronic structure of TMDCs~\cite{Liu2013,Ugeda2014}, in particular the properties of direct excitons, such as their quasiparticle band structure~\cite{Wu2015b,Cheiwchanchamnangij2012b}, binding energies~\cite{Trushin2016,Zhu2015a}, and optical properties~\cite{Klots2014,Hill2015,Wang2015a}.
  As a compliment to the experimental efforts to determine the properties of the energy band structure of charge carriers and the optical properties of excitons in TMDC monolayers, there have been a wide variety of analytical and numerical approaches~\cite{Trushin2016,Hichri2016,Steinhoff2014,Azhikodan2016,Berghauser2014,Kormanyos2013,Berkelbach2013,Komsa2013,Ramasubramaniam2012,Qiu2013,Velizhanin2015} which seek to reproduce experimental results as well as to provide a theoretical framework capable of predicting behavior which has not yet been witnessed.

  TMDC double layer structures have attracted attention for many of the same reasons as monolayer TMDCs, and the scope of experimental and theoretical studies has expanded to fill a variety of niche scenarios which are inaccessible in the case of single TMDC monolayers.
  Many studies of TMDC double layers have focused on indirect excitons, and in particular on their optical properties~\cite{Rigosi2015,Rivera2015,Ceballos2014,Zhu2017,Terrones2013,Yu2015,Huang2014,Liu2014,Hong2014,Wurstbauer2016,Zhu2015a,Zhu2015b,He2014,Amin2015,Debbichi2014,Calman2015,Fang2014,Wang2014} or on the formation and properties of Bose-Einstein condensates and superfluids of indirect excitons~\cite{Fogler2014,Wu2015,Berman2016,Berman2017a}.
  Complexes of indirect excitons, such as trions and biexcitons, can be excited by high intensity light in double layers~\cite{Bondarev2017}.

  In this Paper, we study optical absorption by excitons in TMDCs due to intraexcitonic transitions, that is, transitions from the excitonic ground state to the excited states.
  This is accomplished by solving the Schr\"{o}dinger equation for the electron and hole to obtain the eigenfunctions and eigenenergies of the indirect exciton.
  The obtained solutions allow us to calculate optical quantities such as the oscillator strength, absorption coefficient, and absorption factor, which gives the fraction of incoming photons absorbed by indirect excitons in a TMDC double layer.

  This Paper is organized as follows: in Sec.~\ref{sec:theory}, we present a theoretical description of an indirect exciton formed via the Keldysh potential in a TMDC double layer separated by a dielectric.
  In Sec.~\ref{sec:opticalprops} we derive expressions for the optical properties under consideration.
  A summary of our methodology for obtaining numerical solutions for the electron-hole system with the Keldysh potential follows in Sec.~\ref{sec:methods}.
  We present and discuss all relevant results related to the aforementioned optical properties in Sec.~\ref{sec:results}.
  In Sec.~\ref{sec:comparison}, we compare the optical properties of indirect excitons in a TMDC double layer to the optical properties of direct excitons in monolayer TMDCs, as well as to indirect excitons in coupled quantum wells.
  We discuss our numerical results in the context of experimental research of intraexcitonic optical transitions in semiconductors in Sec.~\ref{sec:future}.
  Our conclusions follow in Sec.~\ref{sec:conclusion}.

\section{\label{sec:theory}The indirect exciton in a TMDC double layer}

  We begin by considering two TMDC monolayers separated by a distance $D$, with many-layer $h$-BN encapsulating the double layer on the top and bottom, and few-layer $h$-BN acting as a dielectric between the two TMDC monolayers.
  The electrons and holes which constitute the indirect excitons are contained in different TMDC monolayers.
  The two-body Schr\"{o}dinger equation for an electron and hole is

\begin{equation}
  \left[ \frac{-\hbar^2}{2} \left( \frac{1}{m_e}\nabla^{2}_{e} + \frac{1}{m_h}\nabla^{2}_{h} \right) + V(\mathbf{r}_e,\mathbf{r}_h) \right]\Psi(\mathbf{r}_e,\mathbf{r}_h) = E\Psi(\mathbf{r}_e,\mathbf{r}_h),
  \label{eq:2bodyschro}
\end{equation}
  where $m_e$ and $m_h$ are the effective electron and hole masses, respectively, $\mathbf{r}_e$ and $\mathbf{r}_h$ are the position vectors for the electron and hole, and $V(\mathbf{r}_e,\mathbf{r}_h)$ is the interaction potential between the electron and hole.
  While the electron and hole interact via the Coulomb potential, in TMDCs the electron-hole interaction is affected by screening which causes the electron-hole attraction to be described by the Keldysh potential~\cite{Keldysh1979}.
  Following the standard procedure for the separation of the relative motion of the electron-hole pair from their center-of-mass motion, one can introduce variables for the center-of-mass of the electron-hole pair, $\mathbf{R}= \frac{m_e \mathbf{r}_e + m_h \mathbf{r}_h}{m_e + m_h}$, and the relative motion of the electron and hole, $\mathbf{r} = \mathbf{r}_e - \mathbf{r}_h$.
 After separation of the electron-hole center-of-mass motion, the Schr\"{o}dinger equation for the relative motion of the electron and hole becomes

\begin{equation}
  \left[ \frac{-\hbar^2}{2 \mu} \nabla^{2}_{\mathbf{r}} + V(\mathbf{r}) \right]\Psi(\mathbf{r}) = E\Psi(\mathbf{r})\label{eq:2bodyschrorel},
\end{equation}
  where $\mu = \frac{m_e m_h}{m_e + m_h}$ is the exciton reduced mass.

  The quantum mechanical properties of the exciton are the subject of this investigation and from this point forward we will examine the eigenfunctions and eigenenergies of the indirect exciton using Eq.~\eqref{eq:2bodyschrorel}.
  We further note that the Keldysh potential has spherical symmetry and only depends on the relative coordinate between the electron and hole.
  Using cylindrical coordinates with the longitudinal axis perpendicular to the planes of the two TMDC monolayers forming a double layer, the relative position vector between the electron and the hole is $\mathbf{r} = \mathbf{r}_e - \mathbf{r}_h = \rho \hat{\bm{\rho}} + D \hat{\mathbf{z}}$, where $\hat{\bm{\rho}}$ and $\hat{\mathbf{z}}$ are unit vectors, $D$ is the fixed interlayer separation, and $\rho$ is the radial separation between the hole and the projection of the electron position onto the TMDC layer with holes.
  In these coordinates the potential $V(r) \equiv V \left( \sqrt{\rho^2 + D^2} \right)$, and the Schr\"{o}dinger equation reads,

\begin{equation}
  \left[ - \frac{\hbar^2}{2 \mu} \left( \frac{\partial^2}{\partial \rho^2} + \frac{1}{\rho} \frac{\partial}{\partial \rho} + \frac{1}{\rho^2}\frac{\partial^2}{\partial\phi^2} \right) + V\left( \sqrt{\rho^2 + D^2} \right) \right] \Psi\left( \rho, \phi \right) = E \Psi\left( \rho, \phi \right).
  \label{eq:expanddelsq}
\end{equation}

Multiplying Eq.~\eqref{eq:expanddelsq} by $\rho^2$ and performing separation of variables $\Psi\left( \rho,\phi \right) = R(\rho)\Phi(\phi)$, we obtain,

\begin{equation}
  \Phi(\phi) = \frac{e^{-i l \phi}}{\sqrt{2 \pi}}, \text{\ \ }l=0,\pm 1,\pm 2,\dots,
  \label{eq:angularsol}
\end{equation}
and $R(\rho)$ is a solution of the following equation:

\begin{equation}
  \frac{d^2 R}{d \rho^2} + \frac{1}{\rho} \frac{d R}{d \rho} + \left[ \frac{2 \mu}{\hbar^2} \left( E - V(\rho) \right) - \frac{l^2}{\rho^2} \right] R(\rho) = 0.
  \label{eq:radialschro}
\end{equation}

The potential $V(\rho)$ is the Keldysh potential~\cite{Keldysh1979}, which in the case of the electron and hole occupying different monolayers is written as

\begin{equation}
  V(\rho) = \frac{\pi k e^2}{2 \kappa \rho_0} \left[ H_{0}\left(\frac{\sqrt{\rho^2 + D^2}}{\rho_0}\right) - Y_{0}\left(\frac{\sqrt{\rho^2 + D^2}}{\rho_0}\right) \right], \label{eq:keldyshpot}
\end{equation}
  where $e$ is the charge of an electron, $k = 9\times 10^{9}~\frac{N m^2}{C^2}$, $\kappa = \frac{\varepsilon_1 + \varepsilon_2}{2}$ describes the surrounding dielectric environment (here $\varepsilon_1$ and $\varepsilon_2$ refer to the dielectric constant of the medium between and surrounding the double layer, respectively), $\rho_0 = \frac{2 \pi \chi_{\text{2D}}}{\kappa}$ is the screening length, where $\chi_{\text{2D}}$ is the 2D polarizability of the medium, and $H_0$ and $Y_0$ are the Struve and Bessel functions of the second kind, respectively.
  Note that $\chi_{\text{2D}}$ is a material property while $\rho_0$ also depends on the environment via the inclusion of $\kappa$ in the denominator.

  We refer to the eigenstates of  for the Keldysh potential by their analogues in the hydrogen atom, that is, $1s$ refers to $(n,l) = (1,0)$, $2s$ is $(n,l) = (2,0)$, $2p$ refers to $(n,l) = (2,\pm 1)$, and so on.

  Eq.~\eqref{eq:radialschro} with the Keldysh potential~\eqref{eq:keldyshpot} is not solvable in closed form.
  It is therefore necessary to turn to numerical methods to solve the Schr\"{o}dinger equation~\eqref{eq:2bodyschrorel} with the Keldysh potential~\eqref{eq:keldyshpot}, allowing us to obtain the eigenfunctions and eigenenergies of the indirect exciton in a TMDC double layer.
  Results of the numerical solutions of the Schr\"{o}dinger equation~\eqref{eq:2bodyschrorel} for an indirect exciton, and the subsequent calculations of the relevant optical properties, are presented in Sec.~\ref{sec:results}.
  % An analytical solution based on an approximate interaction potential is presented in Appendix~\ref{app:theory}.
  % Ultimately, we find that these approximate analytical solutions are a useful tool to verify the accuracy of our numerical solutions, as will be shown in Sec.~\ref{sec:methods}.

\section{\label{sec:opticalprops}Optical absorption by indirect excitons}

To calculate the optical absorption coefficient and related quantities for optical transitions of indirect excitons in TMDC double layers, we need only make slight modifications to the well-established form for the optical absorption due to individual atoms.
%since an exciton is simply a bound state between one positive particle (the hole) and one negative particle (the electron).
Therefore, the expressions that describe the optical properties of indirect excitons have the same functional form as those for the hydrogen atom, even though the dynamics of the underlying eigensystem are quite different.
Additionally, some aspects of this theoretical framework were applied to obtain optical absorption due to magnetoexcitons in semiconductor coupled quantum wells in Ref.~\cite{Lozovik1997}.

  The theoretical treatment of optical absorption by atoms is well-known~\cite{Snoke}.
  Following Ref.~\onlinecite{Snoke}, the oscillator strength, $f_{i \rightarrow f}$, for the transition of the exciton from the initial state $\vert i \rangle$ to the final state $\vert f \rangle$ can be written as:

  %The relevant optical quantities for this Paper are presented below, beginning with the oscillator strength, $f_{i \rightarrow f}$:

  \begin{equation}
	f_{i \rightarrow f} = \left( \frac{2 \mu \omega_{i \rightarrow f} \vert \langle f \vert x \vert i \rangle \vert^2}{\hbar} \right),
	\label{eq:basicf0}
  \end{equation}
  where $\mu$ is the exciton reduced mass, $\omega_{i \rightarrow f} = \left( E_f - E_i \right) / \hbar$ is the Bohr angular frequency of the transition, and $E_i$ and $E_f$ are the eigenenergies of the initial and final states, respectively.
  The oscillator strength, $f_{i \rightarrow f}$, is a dimensionless quantity and obeys the sum rule, $\sum_{f \neq i} f_{i \rightarrow f} = 1$.
  The oscillator strength is interesting from a theoretical viewpoint because it can be used to analyze solely the relative likelihood of a particular system undergoing a particular optical transition.
  Furthermore, calculation of the matrix element $\langle f \vert x \vert i \rangle$ yields the allowed and forbidden transitions.

  For optical transitions in indirect excitons induced by linearly polarized light, the only allowed transitions are to states in which $l_f = l_i \pm 1$, and $n_f \neq n_i$.
  When specifically considering optical transitions from the excitonic ground state $1s$ to the excited states, the only allowed transitions are therefore the excited states $2p$, $3p$, and so on.

  We use $f_0$ to refer to the oscillator strength, and $\omega_0$ to refer to the corresponding Bohr angular frequency, in the cases where the specific states $\vert i \rangle$ and $\vert f \rangle$ under consideration are clear from context, or when speaking generally about the nature of the functions themselves.

  Both the dielectric function $\epsilon(\omega)$ and the electric susceptibility $\chi \left( \omega \right)$, are commonly obtained using \textit{ab initio} techniques~\cite{Kumar2012,Wurstbauer2016,He2014a,Hichri2016,Steinhoff2014,Molina-Sanchez2013,Fogler2014}.
  The imaginary part of the electric susceptibility $\text{Im} \left[ \chi\left( \omega \right) \right]$ is related to the oscillator strength $f_0$ as

  \begin{equation}
	\text{Im}\left[ \chi(\omega) \right] = - \left( \frac{\pi e^2}{2 \varepsilon_0 \mu \omega_0}\frac{n_0}{2h} f_0 \right) \left( \frac{(\Gamma / 2)}{{(\omega_{0}^{2} - \omega^2)}^2 + {(\Gamma / 2)}^2} \right),
	\label{eq:imchi}
  \end{equation}
  where $n_0$ is the 2D concentration of excitons in the TMDC double layer, $h$ is the thickness of one TMDC monolayer, and $\Gamma$ is the damping or homogeneous line-broadening, whose primary physical origin is due to exciton-phonon interactions.
  The fraction $n_0 / (2h)$ represents the 3D concentration of indirect excitons in the TMDC double layer, and the factor of $2$ is included because the single exciton is spread across two TMDC monolayers, each containing either an electron or a hole.
  Unlike the oscillator strength, the imaginary part of the electric susceptibility contains information about the material within which the indirect exciton exists \textendash{} indeed, it may be more accurate to say that the imaginary part of the electric susceptibility is fundamentally a property of the TMDC itself, and a sufficiently thorough calculation of $\text{Im} \left[ \chi(\omega) \right]$ will consider how the TMDC interacts with an incoming photon of any wavelength, and will therefore incorporate the contribution of quasiparticles on the full spectrum $\text{Im} \left[ \chi(\omega) \right]$.
  The presence of quantities such as $n_0$, $h$, and $\Gamma$ demonstrate that $\chi$ is a quantity which depends not only on the specific behavior of the indirect exciton \textendash{} which is encapsulated within $f_0$ \textendash{} but also on material properties such as the thickness of the TMDC monolayer, $h$, the concentration of indirect excitons in the double layer, $n_0$, and the rate at which these excitons interact dissipatively with their surroundings, $\Gamma$.
  It is especially noteworthy that these latter two quantities may be controlled experimentally: $n_0$ by changing the intensity of the pump laser which is creating the indirect excitons, and the damping $\Gamma$ is sensitive to the temperature of the sample, among other things.

  %To the best of the authors' knowledge, published research regarding the maximal, or even typical, 
  Experimental values for the 2D concentration of indirect excitons in TMDC double layers is scarce.
  Recently, $n_0 = 5 \times 10^{15}~\text{m}^{-2}$ was reported for a WSe$_{\text{2}}$ monolayer~\cite{You2015}.
  Below, we are using this experimental value in our calculations, assuming that it is representative of typical concentrations of indirect excitons in a TMDC double layer.

  The second free parameter, $\Gamma$, may in principle be calculated by analyzing an experimentally obtained absorption spectrum \textendash{} in this way, $\Gamma$ is understood as the “line broadening” of each absorption peak, and is thus defined as the full-width half-maximum (FWHM) of each absorption peak.
  Therefore, we must again turn to prior literature to obtain a reasonable value to use for our purposes.
  Optical absorption experiments on indirect excitons in GaAs/GaAlAs quantum wells~\cite{Yuh1988} provide a value on the order of $\Gamma \approx 10^{13}~\text{Hz}$, which corresponds to $41~\text{meV}$.
  Values of the line broadening from recent TMDC optical absorption experiments include for MoS$_{\text{2}}$ phenomenological fits of $30~\text{meV}$~\cite{Mak2010}, $50~\text{meV}$~\cite{Molina-Sanchez2013}, and $20~\text{meV}$~\cite{Steinhoff2014}, which corresponds to $7.2 \times 10^{12}~\text{Hz}$, $1.2 \times 10^{13}~\text{Hz}$, and $4.8 \times 10^{12}~\text{Hz}$, respectively.
  Because $\Gamma$ appears in the denominator of Eq.~\eqref{eq:imchi}, a larger $\Gamma$ corresponds to a smaller maximal value of the absorption, a physically logical result \textendash{} as the damping in an oscillating system grows stronger, its response to the driving force decreases in amplitude.
  Therefore, we use $\Gamma = 10^{13}~\text{Hz}$ throughout our calculations as a conservative approximation of the line broadening in order to avoid overstating the absorption properties of any or all of the TMDC materials studied here.

  The imaginary part of the electric susceptibility is primarily interesting for us because of its close relation to the optical absorption coefficient, $\alpha (\omega)$~\cite{Jackson,HaugnK}:

\begin{equation}
	\alpha(\omega) = - \frac{\omega}{n(\omega) c} \text{Im} \left[ \chi(\omega) \right]. \label{eq:alphachi}
\end{equation}
  In Eq.~\eqref{eq:alphachi}, $n(\omega)$ refers to the refractive index of the environment surrounding the TMDC\@.
  In the case where the environment (here, we consider exclusively $h$-BN) interacts weakly with photons in the frequency range of the corresponding optical transition, we approximate $n(\omega) \approx \sqrt{\varepsilon}$, where $\varepsilon$ is the static dielectric constant of the environment~\cite{HaugnK}, and rewrite Eq.~\eqref{eq:alphachi} as,
\begin{equation}
	\alpha(\omega) = \left( \frac{\omega}{\omega_0 c} \frac{\pi e^2}{2 \varepsilon_0 \sqrt{\varepsilon} \mu}\frac{n_0}{2h} f_0 \right) \left( \frac{(\Gamma / 2)}{{(\omega_{0}^{2} - \omega^2)}^2 + {(\Gamma / 2)}^2} \right). \label{eq:fullalpha}
\end{equation}
% Eq.~\eqref{eq:fullalpha} therefore resembles very closely the expression for the electric susceptibility given in Eq.~\eqref{eq:imchi}, but for the inclusion of the dielectric constant of the $h$-BN, $\epsilon_{h\text{-BN}}=4.89$~\cite{Fogler2014}.
  In much the same way that $f_0$ describes solely how the indirect exciton itself interacts with an incoming photon of a particular frequency, and the imaginary part of the dielectric susceptibility describes how the TMDC material itself interacts with incoming photons of any frequency, the absorption coefficient $\alpha$ further contextualizes the interaction of the photon with the exciton by adding the factor $n(\omega)$ which takes into account the effect of the environment on the exciton-photon interaction.

  It is also important to recognize that Eq.~\eqref{eq:fullalpha} describes the absorption coefficient for a single allowed transition, for example to the $n = 1 \rightarrow 2,~l = 0 \rightarrow 1$ eigenstate.
  However, as was mentioned previously, ground state excitons may transition to two degenerate states, that is, states with $l_f = \pm 1$.
  Therefore, when considering the effect that a photon with angular frequency $\omega_{1 \rightarrow n_f}$ will have as it interacts with the indirect exciton, we must multiply our absorption coefficient $\alpha$ by a factor of two to properly reflect the fact that the photon may induce a transition to either the $l_f = 1$ or $l_f = -1$ eigenstate.

  Using Eqs.~\eqref{eq:imchi} and~\eqref{eq:fullalpha} to obtain the full spectrum of, correspondingly, the imaginary part of the dielectric susceptibility and the absorption coefficient, is now possible.
  To do so, however, one needs a complete theoretical description of how the TMDC in its entirety interacts with an incoming electromagnetic wave of arbitrary frequency, necessarily including knowledge of the band structure of the material as well as any optically accessible phonon modes.

  Instead, let us calculate the maximal value of the absorption coefficient for any given optical transition.
  This maximal value is obtained from Eq.~\eqref{eq:fullalpha} when $\omega = \omega_0$, that is, when the incoming photon's frequency matches exactly with the Bohr angular frequency of the transition in question.
  In this case, the expression for the maximum value of the absorption coefficient, $\alpha(\omega = \omega_0)$, is,
  \begin{equation}
	\alpha(\omega = \omega_0) = \left( \frac{\pi e^2}{2 \sqrt{\varepsilon} \mu \varepsilon_0 c}\frac{n_0}{2h} f_0 \right) \left( \frac{2}{\Gamma} \right).
	\label{eq:alphamax}
  \end{equation}
  Eq.~\eqref{eq:alphamax} is defined by the oscillator strength of the transition, and depends on the following input parameters: the exciton reduced mass, the dielectric constant of the environment, the 2D concentration of indirect excitons, the thickness of the TMDC monolayer, and the value of the line broadening.

  Consider that the absorption coefficient $\alpha$ features prominently in the expression for the intensity of an electromagnetic wave as it propagates a distance $z$ through a homogeneous material, e.g.:

  \begin{equation}
	I \left( z, \omega \right) = I_0 e^{- \alpha(\omega) z},
	\label{eq:IvZ}
  \end{equation}
  where $I_0$ is the initial intensity of the wave.
  In Eq.~\eqref{eq:IvZ}, the physical meaning of the absorption coefficient is clear: it is the inverse of the propagation distance that would correspond to the intensity of the electromagnetic wave decreasing by a factor of $1/e$.
  This form is useful when the electromagnetic wave may propagate any arbitrary distance $z$ through the material described by the absorption coefficient $\alpha$, as is the case in bulk 3D materials, but becomes less useful when the distance $z$ is known and fixed, for example in the case of TMDC double layers.
  Hence, it may be more useful to consider what fraction of the incoming electromagnetic wave is absorbed by a single TMDC double layer system containing a 2D concentration $n_0$ of ground state indirect excitons.

  Let us call this quantity the absorption factor and denote it by $\mathcal{A}$:

  \begin{equation}
	\mathcal{A} \left( \omega=\omega_0 \right) = 1 - \frac{I \left( z=2h, \omega = \omega_0 \right)}{I_0} = 1 - e^{- 2 \alpha(\omega=\omega_0) h}.
	\label{eq:afacdef}
  \end{equation}
  The absorption factor $\mathcal{A}$ simplifies the process of comparing how strongly each TMDC double layer system absorbs incoming light, while taking into account the variety of thicknesses of each TMDC material.

\section{\label{sec:methods}Methodology of numerical calculations}

Let us outline the methodology of numerical calculations for finding the eigenfunctions and eigenenergies of indirect excitons using the Keldysh potential~\cite{Keldysh1979}.
First we consider Eq.~\eqref{eq:2bodyschrorel} for large interlayer separation $D$ and numerically solve the Schr\"{o}dinger equation with the approximate harmonic oscillator potential~\eqref{eq:TexpandV} for different interlayer separations and compare our results to the analytical solutions shown in Eqs.~\eqref{eq:HOeb} and~\eqref{eq:HOef}.
Next, we solve the Schr\"{o}dinger equation with the Keldysh potential for a direct exciton using the same input parameters as in Ref.~\onlinecite{Kylanpaa2015} and compare the solution with results from Ref.~\onlinecite{Kylanpaa2015}.
Finally, the Schr\"{o}dinger equation for the indirect exciton is then solved with the Keldysh potential to obtain the eigenfunctions and eigenenergies of indirect excitons in a TMDC double layer for different values of the interlayer separation $D$.

%%%%%%%%%%%%%%%%%%%%%%%%%%%%%%%%%%%%%%%%%%%%%%%%%

% \begin{table}%[H] add [H] placement to break table across pages
% \caption{\label{}}
% \begin{ruledtabular}
% \begin{tabular}{}
% Lines of table here ending with \\
% \end{tabular}
% \end{ruledtabular}
% \end{table}

\begin{table*}
\caption{\label{tab:matpar}
  Table of relevant material parameters for the calculation of eigenvalues, eigenenergies, and optical properties of various TMDC double layer systems.
For each material, two values of $\mu$ and $\chi_{\text{2D}}$ are given.
The value of the left sub-column corresponds to the value found in the literature which minimizes the indirect exciton binding energy, while the value in the right sub-column maximizes the binding energy.
Because these minimal/maximal values represent the range of values found in the literature, it is highly likely that the true value of each parameter for a given material falls somewhere within the range given, and therefore that the true magnitude of the calculated quantities studied in this paper lies somewhere between the calculated values.
}
\begin{ruledtabular}
\begin{tabular}{ccccccccc}
  Parameter 	& \multicolumn{2}{c}{MoS${}_2$} 	& \multicolumn{2}{c}{MoSe${}_2$}	& \multicolumn{2}{c}{WS${}_2$} 	& \multicolumn{2}{c}{WSe${}_2$} \\ \cline{2-9}
  {}			& $E_{b_{low}}$		&	$E_{b_{high}}$	& $E_{b_{low}}$		&	$E_{b_{high}}$	& $E_{b_{low}}$		&	$E_{b_{high}}$	& $E_{b_{low}}$		&	$E_{b_{high}}$\\ \hline
  $\mu,~[ m_0 ]$ 	& 0.16\cite{Chernikov2014a} &	0.28\cite{Ramasubramaniam2012}%MoS2
					& 0.27\cite{Berkelbach2013} &	0.31\cite{Ramasubramaniam2012}% MoSe2
					& 0.15\cite{Kormanyos2015}	&	0.23\cite{Ramasubramaniam2012}% WS2
					& 0.15\cite{Kormanyos2015}	&	0.27\cite{Ramasubramaniam2012} \\ % WSe2
  $\chi_{\text{2D}},~[ \mbox{\AA} ]$ 	& 7.112\cite{Kylanpaa2015} & 6.60\cite{Berkelbach2013}% MoS2
							& 8.461\cite{Kylanpaa2015} & 8.23\cite{Berkelbach2013}% MoSe2
							& 6.393\cite{Kylanpaa2015} & 6.03\cite{Berkelbach2013}% WS2
							& 7.571\cite{Kylanpaa2015} & 7.18\cite{Berkelbach2013} \\ %WSe2
  $l,~[ \mbox{\AA} ]$\cite{Kylanpaa2015} 		& \multicolumn{2}{c}{6.18} 			& \multicolumn{2}{c}{6.527} 		& \multicolumn{2}{c}{6.219} 	& \multicolumn{2}{c}{6.575} \\
\end{tabular}
\end{ruledtabular}
\end{table*}

  Let us check that our numerical calculations in the case of the harmonic oscillator approximation accurately reproduce the analytical solutions given in Appendix~\ref{app:theory}.
  Verifying the accuracy of the numerical eigenenergies is achieved by straightforwardly comparing the numerically obtained binding energy as a function of $D$ to the binding energy given analytically in Eq.~\eqref{eq:HOeb}.
  Performing such a comparison shows that the numerical values agree with the analytical values to at least four decimal places.

  Verifying the accuracy of the numerical eigenfunctions is best done by comparing the results of a calculated value based on the eigenfunctions themselves.
  To accomplish this, we calculate the in-plane gyration radius of the indirect exciton in the ground state, $r_X = \sqrt{ \langle \rho^2 \rangle} = {\left[ \int \Psi_{1s}^{*} \left( \mathbf{r} \right) \rho^2 \Psi_{1s}\left( \mathbf{r} \right) d\mathbf{r} \right]}^{1/2}$, where $\Psi_{1s}\left( \mathbf{r} \right)$ and $\Psi_{1s}^{*}\left( \mathbf{r} \right)$ represent the ground state excitonic wave function and its complex conjugate, respectively.
  By calculating $r_X$ using both the numerical and theoretical~\eqref{eq:HOef} solutions in the harmonic oscillator approximation presented in Appendix~\ref{app:theory}, we can verify that the numerical calculation produces accurate results.
  Upon comparing these two values, we again find excellent agreement to better than four decimal places.

  To further emphasize that our computational method is sound, we sought to use our computational framework to emulate the results of previously published literature.
  In Ref.~\onlinecite{Kylanpaa2015} the authors used density functional theory to obtain $\mu$ and $\chi_{\text{2D}}$ for MoS$_{\text{2}}$, MoSe$_{\text{2}}$, WS$_{\text{2}}$, and WSe$_{\text{2}}$, and calculated the corresponding binding energy of direct excitons.
  Using the same material parameters from Ref.~\onlinecite{Kylanpaa2015} to calculate the binding energy of direct excitons yields results that agree to better than 1\%.

  Based on the robust agreement between the numerical and analytical results, as well as the robust agreement between our calculations and results from Ref.~\onlinecite{Kylanpaa2015}, there is strong evidence that the code produces accurate eigenfunctions and eigenenergies.
  Given the functional forms of the optical quantities $f_0$, $\alpha$, and $\mathcal{A}$ presented in Sec.~\ref{sec:opticalprops}, we can be assured of accurate calculations of optical absorption provided we are using the correct eigenvalues and eigenfunctions.

  Throughout the previous sections, we have carried out the theoretical analysis of a system consisting of spatially separated electrons and holes under the assumption that the exciton system is dilute enough that it is reasonable to ignore the electrostatic dipole-dipole interaction between the indirect excitons themselves.
  We justify this assumption by comparing the in-plane gyration radius, $r_X$, to the average distance between the excitons themselves, $r_S = 1/\sqrt{\pi n_0}$.
  Rearranging $r_S$, we may write the 2D concentration of indirect excitons in the plane of the TMDC monolayers as $n_0 = 1/\left( \pi r_S^2 \right)$.
  %where now it should be clear that $r_S$ represents the average separation between two indirect excitons, where the circular area $\pi r_S^2$ is the idealized representation of the ``empty space'' between any two neighboring indirect excitons.
  If $r_S \gg r_X$, then the electrostatic interaction between the electron and hole forming an indirect exciton would be much stronger than any dipole-dipole interactions between neighboring excitons.
  It would therefore be reasonable to consider each exciton as effectively isolated from its neighbors, which in turn justifies treating the exciton in Sec.~\ref{sec:theory} as a purely two-body system, and furthermore justifies in Sec.~\ref{sec:opticalprops} the decision to not modify the eigensystem obtained in Sec.~\ref{sec:theory}.

  Using the numerical results for the indirect exciton formed via the Keldysh potential in a TMDC double layer, we find that for $n_0 = 5 \times 10^{15}~\text{m}^{-2}$ the ratio $r_X/r_S$ falls between $0.25$ for $N_{h\text{-BN}}=1$ and $0.5$ for $N_{h\text{-BN}}=9$.
  These results don't necessarily satisfy the condition that $r_S \gg r_X$, nor do they immediately invalidate the assumption that the indirect excitons can be treated as non-interacting.
  It could be argued that as the ratio $r_X/r_S$ approaches $0.5$ at larger interlayer separations that the assumption that the excitons are non-interacting breaks down.
  However, it is worth mentioning again that no experiment yet has used a TMDC double layer with more than five layers of $h$-BN, and $r_X/r_S \approx 0.35 - 0.45$ for $N_{h\text{-BN}} = 5$.
  It is therefore reasonable to assume that our calculated optical quantities, presented in Sec.~\ref{sec:results}, are accurate enough to produce experimentally verifiable results.

\section{\label{sec:results}Results and discussion}

Let us now turn our attention to the calculation of the binding energies, transition energies, oscillator strengths, absorption coefficients, and absorption factors for the $1s \to 2p$ and $1s \to 3p$ transitions using the solution of Eq.~\eqref{eq:radialschro} with the Keldysh potential~\eqref{eq:keldyshpot}.
As the input parameters for calculations of the aforementioned values we use the data listed in Table~\ref{tab:matpar} and the dielectric constant of $h$-BN, $\epsilon_{h\text{-BN}}=4.89$~\cite{Fogler2014}.
With reference to Table~\ref{tab:matpar}, we see that there is still non-trivial disagreement as to the precise value of the material parameters which define each of our TMDC systems.
For that reason, to avoid arbitrarily choosing one value for each parameter and presenting those results as a definitive prediction of the behavior of the systems, we choose the largest and smallest values found in the literature for the material parameters $\mu$ and $\chi_{\text{2D}}$, and use these extreme values to provide upper and lower bounds on each of the quantities we calculate.
Specifically, we find that the largest value of $\mu$ and the smallest value of $\chi_{\text{2D}}$ corresponds to the largest binding energy.
The binding energy is an increasing function of $\mu$ as in the case of the bare Coulomb potential, while the binding energy is a decreasing function of $\chi_{\text{2D}}$.
A small value of $\chi_{\text{2D}}$ corresponds to highly localized screening effects \textendash{} indeed, the Keldysh potential tends exactly to the Coulomb potential in the limit $\chi_{\text{2D}} \to 0$.
Therefore, in the scenario where the TMDC has a higher dielectric constant than the environment, small values of the 2D polarizability $\chi_{\text{2D}}$ correspond to a more Coulomb-like environment,
%where $\varepsilon \rightarrow \varepsilon_{h-\text{BN}}$,
which yields larger binding energies.

\begin{figure}[h!]
  \includegraphics[width=0.75\columnwidth]{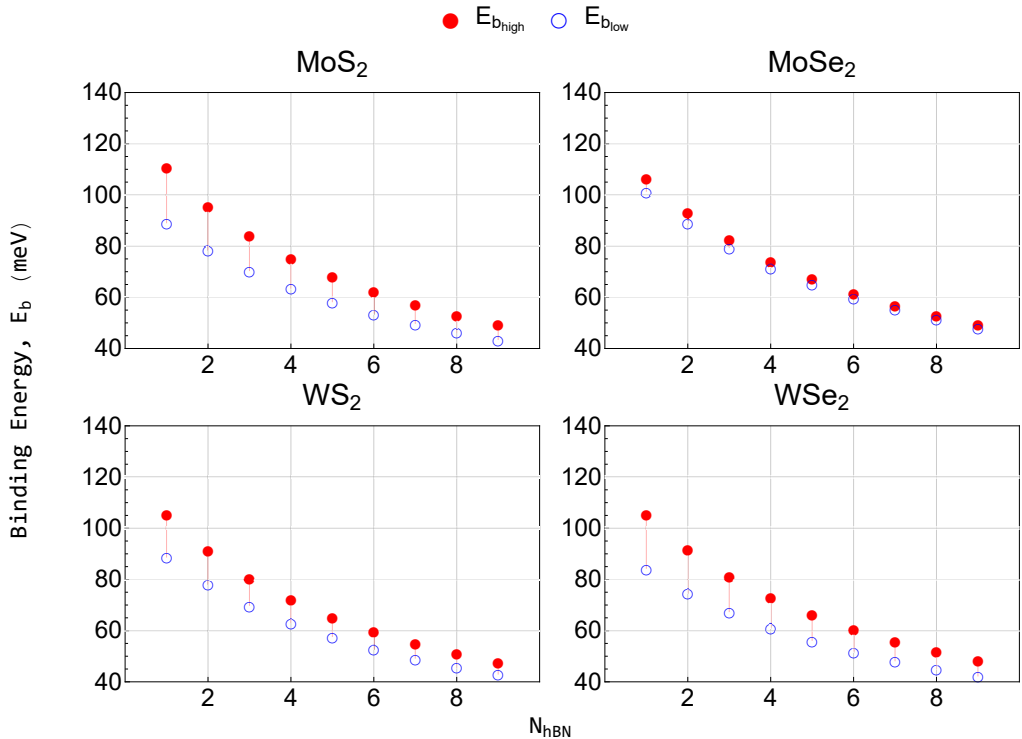}
  \caption{\label{fig:ebrangegrid}Indirect exciton binding energy ranges for the maximal/minimal combinations of material parameters given in Table~\ref{tab:matpar}, plotted as a function of interlayer separation in steps of $D_{hBN} = 0.333$ nm, corresponding to the thickness of one ${h\text{-BN}}$ monolayer.
	The solid circles correspond to the material parameters listed in Table~\ref{tab:matpar} under the $E_{b_{high}}$ subcolumns for each material, while the open circles correspond to the combinations of material parameters listed under the $E_{b_{low}}$ subcolumns.
  This convention is used consistently throughout the paper for all figures.
}
\end{figure}

  Fig.~\ref{fig:ebrangegrid} presents the range of binding energies for the four TMDC materials.
  Results of the calculations show that the binding energy of indirect excitons in each TMDC double layer decreases monotonically as a function of the interlayer separation.
  We see that MoSe${}_2$ has the most tightly constrained upper and lower bounds for the binding energy, which is simply due to MoSe$_{\text{2}}$ having the smallest range of material parameters found in the literature.
  The binding energies for the other three materials are strikingly similar despite the differences in the particular values of $\mu$ and $\chi_{\text{2D}}$ found in Table~\ref{tab:matpar}.
  We also note that the upper and lower bounds of $E_b$ for MoS$_{\text{2}}$, WS$_{\text{2}}$, and WSe$_{\text{2}}$ can differ by nearly 20\% for $N_{h\text{-BN}}=1$, but as the number of $h$-BN monolayers increases to $9$, we see that the difference in binding energy between the upper and lower bounds decreases to roughly 10\%.

  \begin{figure}[h!]
	\centering
	\includegraphics[width=0.75\columnwidth]{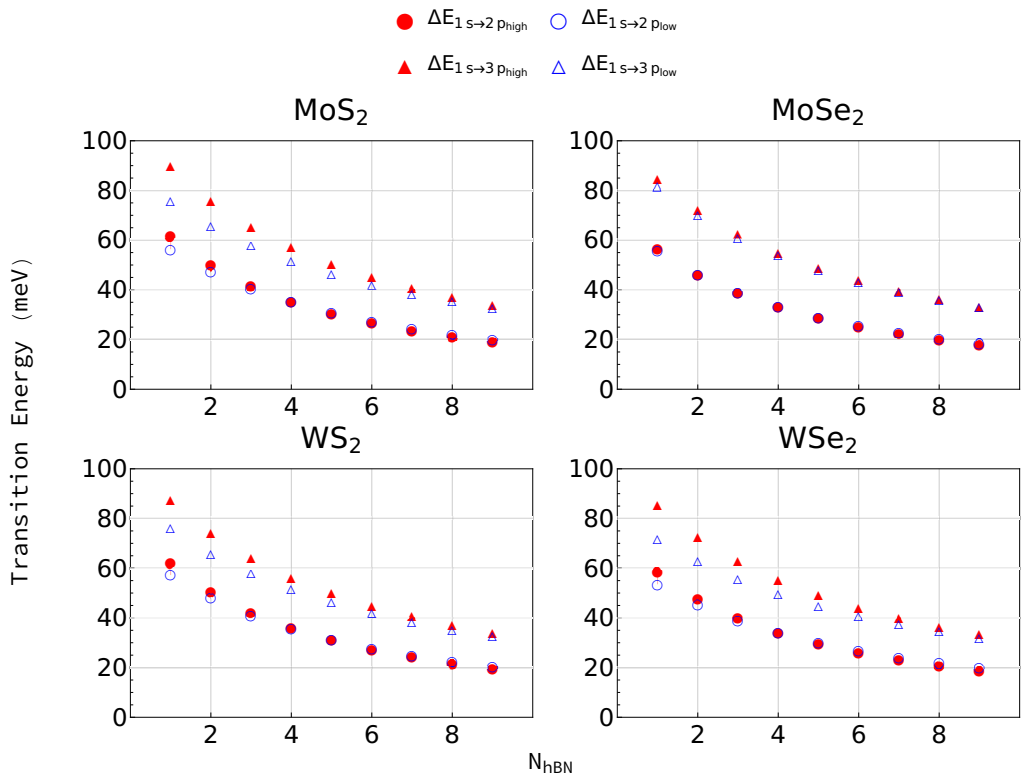}
	\caption{\label{fig:etr}Energy of $1s \to 2p$ (circles) and $1s \to 3p$ (triangles) transitions as a function of the interlayer separation.
	  The solid icons represent the upper bound, while the open icons represent the lower bound.
	}
  \end{figure}

  The energies at which transitions into the excitonic excited states occur are given in Fig.~\ref{fig:etr}.
  It is clear that for both excited states, the transition energy decreases monotonically as the interlayer separation $D$ increases.
  It is notable that there is no overlap between the upper and lower bounds for the $1s \to 2p$ and $1s \to 3p$ transitions for a given value of $N_{h\text{-BN}}$, which suggests that any experimental observation of optical transitions occuring at photon energies given in Fig.~\ref{fig:etr} should unambiguously identify the particular transition which occurred.
  The difference between the upper and lower bounds on the energies of the two transitions shown in Fig.~\ref{fig:etr} is, in general, smaller than the correpsonding difference between the upper and lower bounds on the binding energy as shown in Fig.~\ref{fig:ebrangegrid}.
  This suggests that the values of the eigenenergies themselves are more sensitive to changes in the material parameters than the relative change in energy between successive eigenstates.
  Figs.~\ref{fig:ebrangegrid} and~\ref{fig:etr} also demonstrate that TMDC double layers may be engineered such that the binding energies and transition energies fall within a particular energy range.
  % The transition energies predictably decrease as $D$ increases, however in Fig.~\ref{fig:etr}, each material has a specific value of $D$ at which the upper and lower bounds cross over each other.
  % This is likely the result of the system with the larger binding energy (the ``upper bound'' system) experiencing a larger overall change in its eigenenergies as a function of $D$.
  % Nonetheless, the figures show that the $1s \to 2p$ transition may be accessed by incident photons with energies of around 60 meV in the case of one $h$-BN monolayer, decreasing down to roughly $20$ meV as the number of $h$-BN layers increases to $9$.
  % The $1s \to 3p$ transition requires photon energies in the $75 - 95$ meV range with one $h$-BN monolayer, decreasing to about $30$ meV for all materials when the interlayer $h$-BN increases to $9$ monolayers.

\begin{figure}[h!]
  \centering
  \includegraphics[width=0.75\columnwidth]{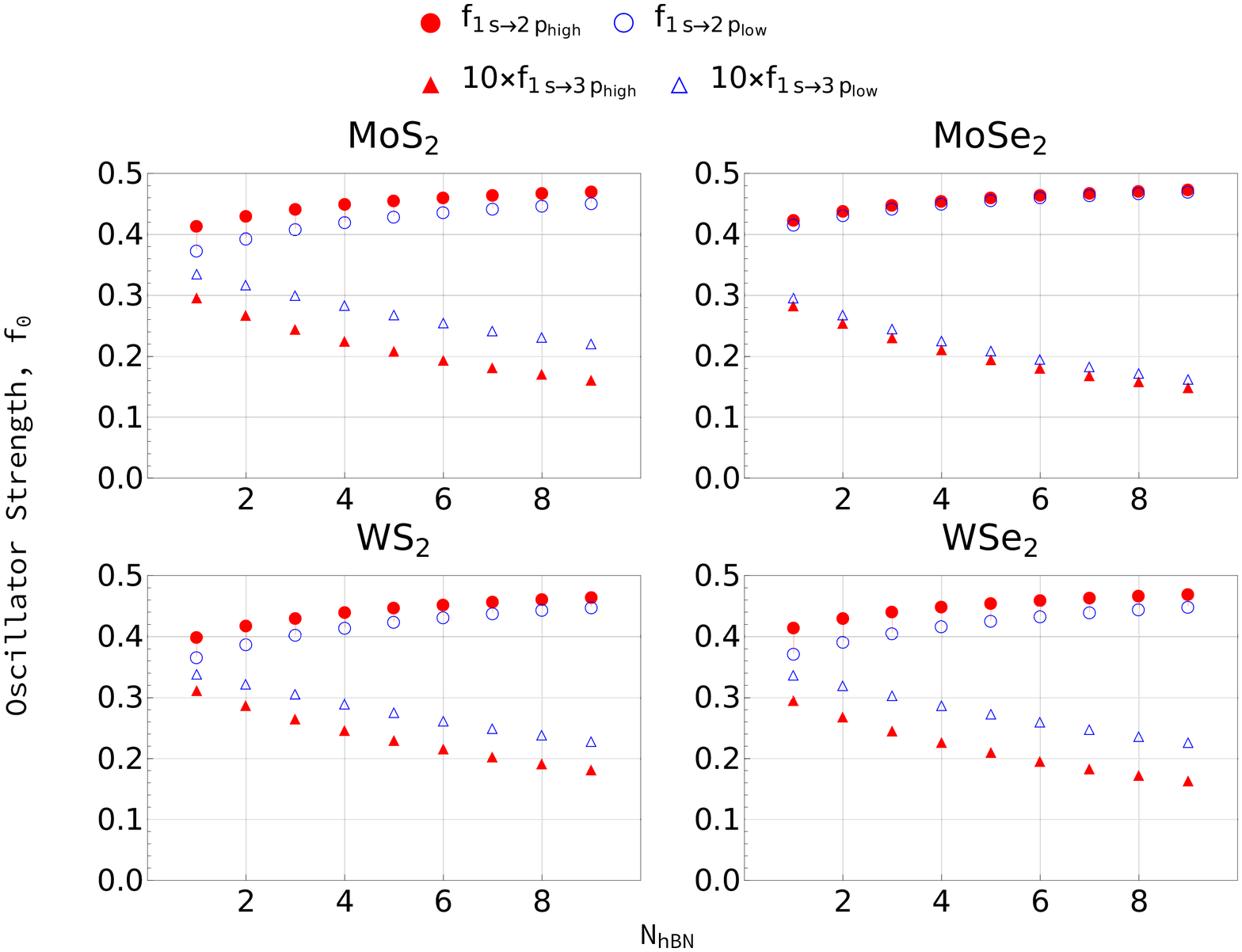}
  \caption{\label{fig:f0}Oscillator strengths for the transition from the $1s$ state to the $2p$ (circles) and $3p$ (triangles) excited states as functions of the interlayer separation. The values of the oscillator strength for the $1s \to 3p$ transition are multiplied by a factor of $10$.}
\end{figure}

Fig.~\ref{fig:f0} presents the dependence of the oscillator strengths for the $1s \rightarrow 2p$ and $1s \rightarrow 3p$ transitions on the interlayer separation.
  The behavior of the oscillator strengths for the two transitions are drastically different: the oscillator strengths for the $1s \to 2p$ transition increase monotonically with increasing $D$, while the oscillator strengths for the $1s \to 3p$ transition decrease monotonically.
  Moreover, the oscillator strengths for the $1s \to 2p$ transition are roughly an order of magnitude larger than for the $1s \to 3p$ transition.
  The upper and lower bounds of the oscillator strengths for a given transition and interlayer separation show much less variability overall than either the binding energies or transition energies, however the difference between the upper and lower bounds remains relatively constant as the interlayer distance increases.
  This may be due in part to the fact that the sum of the oscillator strengths for all allowed transitions must be unity, fundamentally constraining the range of possible values.

  Now that the results for the oscillator strength $f_0$ have been presented, we turn our attention again to Eq.~\eqref{eq:alphamax}, and recognize that we may rewrite the expression for $\alpha$ as:

  \begin{equation}
	\alpha = C_{\alpha_{i}} f_0, \label{}
	\label{eq:alphascale}
  \end{equation}
  where

  \begin{equation}
	C_{\alpha_{i}} = \frac{\pi e^2}{2 \epsilon_0 \sqrt{\epsilon} c \mu_{i}} \frac{n_0}{2h} \frac{2}{\Gamma}
	\label{eq:scaledef}
  \end{equation}
is the absorption coefficient scale factor and $i = \text{MoS}_{2},~\text{MoSe}_{2},~\text{WS}_{2},~\text{WSe}_{2}$.
 Hence, one may simply use the value of $f_0$ found in Fig.~\ref{fig:f0} paired with the appropriate scale factor given in Table~\ref{tab:scale} to obtain $\alpha$.
 The absorption factor $\mathcal{A}$ may then be straightforwardly calculated using Eq.~\eqref{eq:afacdef}.

\bgroup%
  \def\arraystretch{1.25}
\begin{table*}
\caption{\label{tab:scale}
  Table of the absorption coefficient scale factor, $C_{\alpha}$, for each combination of material parameters as outlined in Table~\ref{tab:matpar}.
  Multiplying the appropriate value of the scale factor by the corresponding oscillator strength $f_0$ yields the absorption coefficient $\alpha$ in units of $\left[ 10^6~\text{m}^{-1} \right]$.
  One may then obtain the absorption factor $\mathcal{A}$ by taking the absorption coefficient, calculated using the scale factor below, and plugging it into Eq.~\eqref{eq:alphascale}.
}
  \begin{ruledtabular}
	\begin{tabular}{ccccccccc}
  Parameter 	& \multicolumn{2}{c}{MoS${}_2$} 	& \multicolumn{2}{c}{MoSe${}_2$}	& \multicolumn{2}{c}{WS${}_2$} 	& \multicolumn{2}{c}{WSe${}_2$} \\ \cline{2-9}
  {}			& $E_{b_{low}}$		&	$E_{b_{high}}$	& $E_{b_{low}}$		&	$E_{b_{high}}$	& $E_{b_{low}}$		&	$E_{b_{high}}$	& $E_{b_{low}}$		&	$E_{b_{high}}$\\ \hline
  $C_{\alpha}$, $\left[ 10^6~\text{m}^{-1} \right]$ 	& 82.29	&	144.01	%MoS2
											& 70.38	&	80.80	%MoSe2
											& 99.55	&	152.65	%WS2
											& 80.21	&	144.38 \\ % WSe2
	\end{tabular}
  \end{ruledtabular}
\end{table*}
\egroup%

  Though our calculations show that MoSe${}_2$ has the largest oscillator strength, it actually has the smallest absorption coefficient of the four TMDC materials.
  We also find that WS${}_2$ has the largest absorption coefficient, with MoS${}_2$ and WSe${}_2$ being roughly equal.
  Since $\mathcal{A}$ is very closely related to the absorption coefficient $\alpha$, we draw many of the same conclusions as for $\alpha$.
  For the $1s \rightarrow 2p$ transition, WS${}_2$ and WSe${}_2$ absorb the most strongly, showing between $2.0\%$ and $3.5\%$ absorption in the case of one $h$-BN monolayer, increasing to between $2.5\%$ and $4.1\%$ absorption when $D$ increases to $9$ layers of $h$-BN\@.
  MoSe${}_2$ again exhibits the weakest and least variable absorption, barely surpassing an upper bound of $2.5\%$ absorption for $9$ layers of $h$-BN\@.
  In general, the absorption coefficients and absorption factors for the $1s \to 3p$ transition are about an order of magnitude less than the corresponding values for the $1s \to 2p$ transition.

\section{\label{sec:comparison}Comparison of results to optical absorption in TMDC monolayers and coupled quantum wells}

  Despite the fundamental differences between interband optical absorption in TMDC monolayers and the intraexcitonic optical absorption studied here, it may still prove instructive to examine the underlying physical differences between the two processes.
  In light of the fundamental differences between interband and intraexcitonic optical absorption, it may be helpful to place our results in the broader context of optical absorption in TMDCs.
  It was previously reported~\cite{Xia2014,Eda2013,Mak2010,Wang2014} that interband absorption leading to the creation of direct excitons in monolayer MoS${}_2$ is nearly $10\%$, e.g. $\mathcal{A} = 0.10$.
  The creation of indirect excitons in TMDC double layers, on the other hand, is a multi-step process \textendash{} first, \textit{direct} excitons are created in a TMDC monolayer via optical excitation in the presence of an electric field perpendicular to the plane of the monolayers.
  The electric field separates the electron and hole into the spatially separated TMDC monolayers of the double layer system, at which point the excited state transitions studied here may be accessed optically.
  Therefore, despite the fact that the aforementioned studies on interband transitions leading to the creation of direct excitons focused specifically on TMDC monolayers, the full absorption spectrum for TMDC double layer systems will include both the full absorption spectrum for direct-exciton-creating interband transitions plus the absorption spectrum corresponding to the transitions to indirect excitonic excited states considered in this Paper.

  %%%% Got rid of this paragraph because it's too off-topic (might be nice for thesis though)
%   We furthermore note that where $\alpha \approx 10^{7}~\text{m}^{-1}$ and the thickness of a TMDC double layer $l \approx 6 \mbox{\AA}$, the argument of the exponential function, $(- 2 \alpha l)$, is extremely small.
% It is therefore valid to approximate the exponential term in Eq.~\eqref{eq:afacdef} as a linear function by taking a first order Taylor expansion, and we obtain $\mathcal{A} \approx 1 - \left( 1 - (2 \alpha l) \right) = 2 \alpha l$.
% Therefore, to a very good approximation, the absorption factor $\mathcal{A}$ may be straightforwardly calculated by taking the absorption coefficient $\alpha$ and multiplying by $2l$, the thickness of a TMDC double layer.
%  Using the scale factors of Table~\ref{tab:scale}, there is a percent error of between $0.1\%$ and $0.3\%$ when using $\mathcal{A} \approx 2 \alpha l$ as opposed to the full expression given by Eq.~\eqref{eq:afacdef}.

  Semiconductor coupled quantum wells (CQWs) such as GaAs/GaAlAs have enjoyed intensive experimental~\cite{Masselink1985,Damen1990,Miller1981,Fox1991} and theoretical~\cite{Chuang1991,Yuh1988,Schmitt-Rink1985,Andreani1990,Sivalertporn2012} study.
  Experimental results for the optical absorption coefficient for interband optical transitions in CQWs, when not presented in arbitrary units, show that the first interband optical transition~\cite{Masselink1985} has an absorption coefficient of around $4 \times 10^6~\text{m}^{-1}$.
  This value of $\alpha$ for interband transitions is therefore roughly an order of magnitude smaller than our current results for intraexcitonic transitions.
  TMDCs are a topic of intense study precisely because they exhibit exceptionally strong absorption.
  In comparison to quantum wells, we see that the absorption coefficients $\alpha$ for interband optical transitions in TMDCs are roughly an order of magnitude greater than the absorption coefficients $\alpha$ for the corresponding transitions in semiconductor coupled quantum wells.
  % Thus, interband transitions in TMDCs are one order of magnitude higher than our intraexcitonic results, another logically satisfactory conclusion.

\section{\label{sec:future}Relation to experiment}

  Intraexcitonic optical transitions have been studied experimentally since at least 2004, when such a study was performed in Cu${}_{2}$O~\cite{Kuwata-Gonokami2004,Kubouchi2005,Jorger2005,Huber2006}.
  Additional experiments were performed in GaAs/GaAlAs quantum wells~\cite{Huber2005,Huber2008}.
  More recently, similar experiments were performed in monolayer TMDCs~\cite{Poellmann2015b,Cha2016,Steinleitner2017} in which the excited states of the direct exciton were probed.
  However, it appears that these types of studies have not yet been performed for indirect excitons in TMDC double layer systems.
  Therefore, while we are unable to directly compare our results with previously published literature, we state with confidence that our theoretical approach is sound, and that it is certainly possible to design an experiment which could probe exactly the types of optical transitions that are studied here.

  Though our computational results are limited by the accuracy of the input parameters shown in Table~\ref{tab:matpar}, this general disagreement as to the precise values of the material properties of TMDCs does not invalidate the predictive power of our calculated values of the binding energy, transition energies, oscillator strengths, absorption coefficients, and absorption factors for indirect excitons in TMDC double layers.
  On the contrary, by surveying the range of possible values for experimentally verifiable quantities ranging from the binding energies to the absorption coefficients, we provide here convenient upper and lower bounds for all optically relevant quantities for the benefit of future experimentalists who seek to observe the phenomena predicted here.
  In principle, the transition energies given in Fig.~\ref{fig:etr} may be verified experimentally using two-photon spectroscopy, using an experimental procedure that resembles those found in the Papers cited in the previous paragraph.
  Direct observation of absorption or photoluminescence at photon frequencies corresponding to the transition energies presented in Fig.~\ref{fig:etr} should be an unambiguous confirmation of predicted transitions, but also a demonstration of the feasibility of taking advantage of these types of optical transitions for future optoelectronic applications.

\section{\label{sec:conclusion}Conclusions}

  To conclude, we provide experimentally verifiable predictions for the binding energies, and the transition energies, oscillator strengths, absorption coefficients, and absorption factors for the $1s \to 2p$ and $1s \to 3p$ transitions.
  Using solutions of the Schr\"{o}dinger equation with the Keldysh potential for an indirect exciton in a TMDC double layer, we calculated binding energies, transition energies, oscillator strengths, absorption coefficients, and absorption factors and their dependence on interlayer separation for four different TMDC materials.

  We emphasize the following conclusions, which are generally applicable to each of the TMDC materials studied here:
  i) The binding and transition energies for an indirect exciton in double layer TMDC semiconductors monotonically decrease as the interlayer separation is increased: binding energies decrease by nearly a factor of two, while energies for the $1s \to 2p$ and $1s \to 3p$ transitions decrease by nearly a factor of three as the number of $h$-BN monolayers is increased from $1$ to $9$.
  ii) The oscillator strengths for the $1s \to 2p$ transition increase monotonically with increasing $D$, while the oscillator strengths for the $1s \to 3p$ transition decrease monotonically.
  Moreover, the oscillator strengths for the $1s \to 2p$ transition are roughly an order of magnitude larger than the $1s \to 3p$ transition for a single $h$-BN monolayer, becoming about $20$ times larger when the interlayer separation is increased to $9$ layers of $h$-BN.
  iii) The absorption coefficients and absorption factors show that indirect excitons in TMDC double layers should exhibit exceptionally strong absorption.

While our calculations are limited by the range of values of the input parameters presented in Table~\ref{tab:matpar}, our predicted values for the binding and transition energies, oscillator strengths, absorption coefficients, and absorption factors are, in principle, accurate enough to be verified experimentally.
In fact, the deliberate decision to present these calculations as upper and lower bounds on observable quantities is made specifically to aid experimentalists who seek to study the optical properties of indirect excitons.
We hope that our results provide a roadmap to enable experimentalists to verify that indirect excitons have been created in a TMDC double layer system.

\appendix
\section{\label{app:theory}The harmonic oscillator approximation of the Keldysh potential}

We emphasize that the purpose of the following derivation is to provide an analytical frame of reference for the numerical solutions of the Schr\"{o}dinger equation~\eqref{eq:2bodyschrorel} with the Keldysh potential~\eqref{eq:keldyshpot}.
  The Schr\"{o}dinger equation, Eq.~\eqref{eq:2bodyschrorel}, with the Keldysh potential, Eq.~\eqref{eq:keldyshpot}, is not solvable in closed form.
  However, in the case when the interlayer separation $D$ is much larger than the in-plane gyration radius, $r_X = \sqrt{\langle \rho^2 \rangle}$, we may perform a Taylor expansion of Eq.~\eqref{eq:keldyshpot} about $\rho=0$ up to first order with respect to ${(\rho/D)}^2$ and obtain:

\begin{equation}
	V(\rho) \approx -V_0 + \gamma^2 \rho^2,
	\label{eq:TexpandV}
\end{equation}
where

\begin{subequations}\label{eq:HOterms}
\begin{eqnarray}
  V_0 = & \frac{\pi k e^2}{2 \kappa \rho_0} \left[ H_0 \left( \frac{D}{\rho_0} \right) - Y_0 \left( \frac{D}{\rho_0} \right) \right], \label{eq:HOtermsA} \\
  \gamma^2 = & - \frac{\pi k e^2}{4 \kappa \rho_{0}^{2} D} \left[ H_{-1} \left( \frac{D}{\rho_0} \right) - Y_{-1} \left( \frac{D}{\rho_0} \right) \right]. \label{eq:HOtermsB}
\end{eqnarray}
\end{subequations}
In Eq.~\eqref{eq:HOtermsB}, $H_{-1}$ and $Y_{-1}$ are the Struve and Bessel functions of the second kind, of order $\nu = -1$.
Having recast the potential in the form of the quantum harmonic oscillator, we now have a Schr\"{o}dinger equation which has an analytical solution~\cite{cotan2} in terms of the associated Laguerre polynomials $L_{n}^{l}$:

\begin{equation}
  \Psi(\rho) = {(-1)}^{\frac{n - \lvert l \rvert}{2}} {\left[ \frac{\left( \frac{n - \lvert l \rvert}{2} \right)!}{\left( \frac{n + \lvert l \rvert}{2} \right)!} \right]}^{\frac{1}{2}} (\gamma) e^{\frac{- \gamma^2 \rho^2}{2}} {(\gamma^2 \rho^2)}^{\frac{\lvert l \rvert}{2}} L^{\lvert l \rvert}_{\frac{n - \lvert l \rvert}{2}} (\gamma^2 \rho^2) \frac{e^{i l \varphi}}{\sqrt{2 \pi}},\label{eq:HOef}
\end{equation}
where $L_{n}^{l}$ is the associated Laguerre polynomial of degree $n$.
The corresponding eigenenergies are:

\begin{equation}
  E_{n} = {\left[ \frac{2 \hbar^2 \gamma^2}{\mu} \right]}^{1/2} \left( n +1 \right) - V_0, \\
  \label{eq:HOeb}
\end{equation}

where $V_0$ and $\gamma^2$ are defined by Eqs.~\eqref{eq:HOtermsA} and~\eqref{eq:HOtermsB}, respectively and the quantum numbers $n=0,1,2,\dots$ and $l=-n,-n+2,\dots,n$ are the principal and angular momentum quantum numbers, respectively.

\bibliography{paper-1}

\end{document}